\documentclass[12pt,onecolumn,floatfix,altaffilletter,superscriptaddress,
               tightenlines,showpacs,showkeys,preprintnumbers,nofootinbib]{revtex4-1}
\pdfoutput=1
\usepackage[colorlinks=true,citecolor=blue,linkcolor=blue,breaklinks=true]{hyperref}
\usepackage{amsmath,amssymb}
\usepackage{epsfig}  
\usepackage{graphicx}                
\usepackage{url}
\usepackage{color}
\usepackage{multirow}
\usepackage{placeins}
\usepackage[dvipsnames]{xcolor}
\allowdisplaybreaks

\bibpunct{[}{]}{,}{n}{}{,}

\renewcommand{\vec}[1]{{\mathbf{#1}}}

\newcommand{\w}{\omega}
\newcommand{\lam}{\lambda}

\begin{document}

\title{Fast neutrino flavor conversions near the supernova core with \\realistic flavor-dependent angular distributions}

\author{Basudeb Dasgupta}
\email{bdasgupta@theory.tifr.res.in}
\affiliation{Tata Institute of Fundamental Research,
             Homi Bhabha Road, Mumbai, 400005, India.}
\author{Alessandro Mirizzi}
\email{alessandro.mirizzi@ba.infn.it }
\affiliation{Dipartimento Interateneo di Fisica ``Michelangelo Merlin'', Via Amendola 173, 70126 Bari, Italy.}
\affiliation{Istituto Nazionale di Fisica Nucleare - Sezione di Bari,
Via Amendola 173, 70126 Bari, Italy.}
\author{Manibrata Sen}
\email{manibrata.sen@gmail.com }  
\affiliation{Tata Institute of Fundamental Research,
             Homi Bhabha Road, Mumbai, 400005, India.}
\preprint{TIFR/TH/16-35}
\date{September 9, 2016}

\begin{abstract}
It has been recently pointed out that neutrino fluxes from a supernova 
can show substantial flavor conversions almost immediately above the 
core. Using linear stability analyses and numerical solutions of the 
fully nonlinear equations of motion, we perform a detailed study of these 
\emph{fast conversions}, 
focussing on the region just above the supernova core. 
We carefully specify the instabilities for evolution in space or time, and
find that neutrinos travelling towards the core make fast conversions 
more generic, i.e., possible for a wider range of flux ratios and angular asymmetries that produce
a \emph{crossing} between the zenith-angle spectra of $\nu_e$ and ${\bar\nu_e}$.
Using fluxes and angular distributions 
predicted by supernova simulations, we find that fast conversions can 
occur within tens of nanoseconds, only a few meters away from the putative 
neutrinospheres.  If these fast flavor conversions indeed take place, they would have important 
implications for the supernova explosion mechanism and nucleosynthesis.
\end{abstract}

\maketitle

\section{Introduction}
\label{sec:1}

Flavor conversions of supernova (SN) neutrinos have been a field of intense research, with hundreds of studies aimed at shedding light on the fascinating physics of
 neutrino flavor changes during the gravitational collapse of a massive star (see \cite{Mirizzi:2015eza} for a recent review).  The early studies in this field considered only the impact of vacuum mixing and the Mikheyev-Smirnov-Wolfenstein (MSW) matter effect \cite{Wolfenstein:1977ue,Mikheev:1986gs}. Within this \emph{MSW paradigm} of SN neutrino oscillations~\cite{Dighe:1999bi}, large flavor conversions were thought to be possible when the neutrino vacuum oscillation frequency $\omega= \Delta m^2/2E$, associated with one of the two neutrino mass-squared differences $\Delta m^2$, is of the order of the matter potential  $\lambda= \sqrt{2}G_F n_e$, associated with the net electron density.
For typical dynamical SN post-bounce matter profiles this condition is fulfilled at radii $r \sim {\mathcal O} (10^3)$\,km from the SN core, where $\omega\simeq\lambda$, and was expected to be the primary mechanism for flavor conversion. The sensitivity of these matter effects to shock-wave propagation \cite{Schirato:2002tg,Fogli:2003dw,Fogli:2004ff,Tomas:2004gr} and to fluctuations in the density of the stellar envelope \cite{Dasgupta:2005wn, Borriello:2013tha} was also thoroughly investigated.

In the deepest SN regions, the neutrino density itself is so high that an additional flavor-non-diagonal $\nu$-$\nu$ potential, $\mu \sim \sqrt{2} G_F n_\nu$, exists for the neutrino propagation~\cite{Pantaleone:1992eq}. This potential induces \emph{collective} or  \emph{self-induced} flavor oscillations with a frequency $\sim \sqrt{\omega \mu}$. A decade ago, based on improved numerical calculations and theoretical understanding, it was established that the MSW effect is indeed not the only route for large neutrino flavor conversions in a SN~\cite{Duan:2006an,Hannestad:2006nj} (see refs.~\cite{Duan:2010bg,Mirizzi:2015eza} for details). These studies predicted large collective flavor conversions at $r \sim {\mathcal O} (10^2)$\,km, where $\omega\simeq\mu$. Broadly, these self-induced flavor conversions are expected to either produce smeared-out spectral swaps in the neutrino fluxes \cite{Fogli:2007bk,Dasgupta:2009mg} or flavor decoherence \cite{EstebanPretel:2007ec} leading to a flavor  equilibration among the different species. In detail however, the rich and surprising phenomenology of collective oscillations remains a subject of active research and we are still far from having a coherent picture. See ref.~\cite{Chakraborty:2016yeg} for a discussion of the recent developments and the open issues.

R. Saywer had pointed out already in 2005 that the $\nu$-$\nu$ potential would lead to even faster flavor conversions at a rate $\sim\mu$, in contrast to $\sqrt{\omega \mu}$ above, producing flavor equilibrium among the different neutrino fluxes at $r \sim {\mathcal O} (1)$\,m from the SN surface~\cite{Sawyer:2005jk}.
Surprisingly, these conversions were predicted to occur even for (almost) massless neutrinos, requiring a nonzero $\omega$ perhaps only as an initial disturbance. Therefore, these conversions would be independent of the yet unknown neutrino mass ordering.
It was stated that the necessary condition to achieve these fast conversions is the presence of significantly different angular distributions for the different neutrino flavors. Indeed, since non-electron flavors
 $\nu_x = {\bar\nu}_x\equiv\nu_{\mu,\tau}$ decouple from matter deeper than ${\bar\nu_e}$, and these latter deeper than ${\nu_e}$, one expects that close to
the SN core the  $\nu_x$ zenith-angle distribution would be more forward-peaked than that of ${\bar\nu_e}$, which in turn would be more forward-peaked than the $\nu_e$ distribution. The latter, in particular, would contain a significant fraction of inward-going neutrinos.

This study was developed in a following paper \cite{Sawyer:2008zs} by the same author, where a stability analysis of the linearized equations of motion showed the presence of rapidly growing run-away modes, leading to a speed-up of the flavor conversions in the presence of non-trivial
$\nu$ angular distributions. The author explicitly noted that these effects were not found in other numerical studies of self-induced flavor conversion because these 
latter focussed on a region relatively far away from the neutrinosphere, where the angular distributions of the different $\nu$ species become rather similar.

These remarkable findings presented in refs.~\cite{Sawyer:2005jk,Sawyer:2008zs} were obtained using very schematic models. Specifically, the calculations were performed with highly discretized versions of the angular distributions, with very few modes. Drastic discretization of angular distributions were known to lead to spurious flavor conversions that disappear when a sufficiently large number of angular bins are
 adopted~\cite{Sarikas:2012ad}. 
Moreover, close to the neutrinosphere, it was expected that the matter potential would be much larger than the neutrino potential. This would prevent self-induced flavor conversions through the  ``multi-angle matter suppression'' mechanism~\cite{EstebanPretel:2008ni,Chakraborty:2011nf, Chakraborty:2011gd}. These caveats may have discouraged further investigation of this issue.

Recently~\cite{Sawyer:2015dsa}, Sawyer presented new results with a stability analysis  obtained using a larger number of angular bins. Fast conversions were shown to survive, while the spurious instabilities get suppressed by increasing the number of modes.  Moreover, he clarified that even if the large matter term would suppress the flavor conversions \emph{in space}, if one relaxes the assumption of stationarity of the solution, very fast turnover \emph{in time} would occur close to the SN core.

Fast conversions were further studied in \cite{Chakraborty:2016lct}, where a stability analysis with non-trivial \emph{continuous} $\nu$ angular distributions was performed, at large distances from the core. Here, it was found that fast conversions typically take place for physically less plausible conditions, i.e., if there are more $\bar\nu_e$ than $\nu_e$, or if the emission distribution of $\bar\nu_e$ are wider, unlike what was found by Sawyer, except if $\mu\simeq\lambda$. Another puzzling aspect was why the instabilities found here, for physically plausible neutrino fluxes and $\mu\simeq\lambda$, did not seem to require any departure from exact azimuthal symmetry, while Sawyer found fast conversions even when ordinary matter in the background was ignored, as long as there was some noise in the emission distributions. As we will show later, the resolution to these differences appears to be that Sawyer's studies have focussed on fast flavor evolution in \emph{time}, very close to the SN core using discrete distributions, whereas, the analysis in ref.~\cite{Chakraborty:2016lct} focusses on the flavor evolution in \emph{space}, at larger distances and employing continuous distributions.

The purpose of this paper is a more detailed study of the conditions for the development of the fast flavor conversions close to the SN core. In particular, we revisit the results of refs.~\cite{Sawyer:2005jk,Sawyer:2008zs,Sawyer:2015dsa,Chakraborty:2016lct}, and redo the linear stability analyses for a flat source geometry that more appropriately models the SN emission region, focussing on physically well-motivated neutrino fluxes, {i.e., $\nu_e$ have a larger flux and  wider angular distribution than $\bar\nu_e$}, and using continuous distributions to avoid the problem of spurious modes. We carefully specify the possible instabilities for evolution in space or time, and show the impact of inward-going neutrinos. We verify these linear stability predictions, for several cases, using numerical calculations of the fully nonlinear evolution. Finally, we also present fully nonlinear calculations of fast conversion using SN neutrino fluxes and angular distributions from a high quality SN simulation.

The paper is organised as follows. In Sec.\,\ref{sec:2} we set up the non-linear equations of motion for flavor evolution of neutrinos emitted from a flat source, modelling the SN emission region. In Sec.\,\ref{sec:3} we present the results of linear stability analyses of the flavor evolution, employing zenith angle spectra of $\nu_e$ and $\bar\nu_e$ that show a \emph{crossing}. We systematically study the possibility for obtaining fast conversions for evolution in space or time, showing that background matter suppresses instabilities in space evolution but not in time. We show that neutrinos travelling inwards make fast conversion even more generic. In Sec.\,\ref{sec:4}, we present numerical results using neutrino angular distributions extracted from a high-quality SN simulation, showing that fast conversions may occur with realistic neutrino fluxes. In Sec.\,\ref{sec:5}, we conclude.

\section{Set-up of the flavor evolution} 
\label{sec:2}

In absence  of external forces acting on neutrinos, the dynamics
of the  space-dependent $\nu$ occupation numbers or Wigner function $\varrho_{{\bf p}, {\bf x},t}$
with momentum ${\bf p}$ at position ${\bf x}$ and time $t$ is dictated by the following equations of motion (EoMs)~\cite{Sigl:1992fn,Strack:2005ux} 

\begin{equation}
\partial_t \varrho_{{\bf p}, {\bf x},t} + {\bf v}_{\bf p} \cdot \nabla_{\bf x} \varrho_{{\bf p}, {\bf x},t} 
= - i [\Omega_{{\bf p}, {\bf x},t}, \varrho_{{\bf p}, {\bf x},t}] +{\mathcal C}[ \varrho_{{\bf p}, {\bf x},t}]
\,\ ,
\label{eq:eom}
\end{equation}
with the Liouville operator on the left-hand side. To lighten the notation, we shall drop the subscripts ${\bf x}$ and $t$, which emphasize that the density matrices and the Hamiltonian are dependent on space and time. The first term in Eq.\,(\ref{eq:eom}) accounts for explicit time
dependence, while the second term, proportional to the neutrino velocity ${\bf v_p}$, encodes the spatial dependence due to particle free streaming. On the right-hand-side, the matrix $\Omega_{{\bf p}}$ is the Hamiltonian
\begin{equation}
\Omega_{{\bf p}}= \Omega_{{\rm vac}} + \Omega_{\rm MSW} + \Omega_{\nu\nu} \,\ ,
\label{eq:ham}
\end{equation}
containing the vacuum, matter and self-interaction terms, that leads to the evolution of $\varrho_{{\bf p}}$ over space and time.
 
Simplifying to an effective two-flavor scenario, the matrix of vacuum oscillation frequency is 
$\Omega_{{\rm vac}}= \textrm{diag}(-\omega/2, +\omega/2)$ 
in the mass basis, where $\omega=(m_2^2-m_1^2)/(2E)$, where $E=|{\bf p}|$ for 
ultra-relativistic neutrinos.  For antineutrinos, the EoMs are the same but with the replacement $\Omega_{{\rm vac}} \to - \Omega_{{\rm vac}}$, thus it is convenient to think of antineutrinos of energy $E$ as neutrinos of energy $-E$, making their EoMs identical. The matter effect in Eq.\,(\ref{eq:ham}), due to background 
electron density $n_e$, is represented by
\begin{equation}
\Omega_{\rm MSW}=  \lambda\,\ \textrm{diag} (1,0) \,\ ,
\end{equation}
in the weak interaction basis, where $\lambda =\sqrt{2} G_F n_e$. 
Finally, the effective Hamiltonian due to $\nu-\nu$ interactions is given by
\begin{equation}
\Omega_{\nu\nu} = \sqrt{2} G_F \int \frac{d^3 {\bf q}}{(2 \pi)^3} ({\varrho_{\bf q}} - {\bar\varrho_{\bf q}}) (1 -{\bf v}_{\bf p}\cdot {\bf v}_{\bf q}) \,\ ,
\end{equation}
where the term $(1 -{\bf v}_{\bf p}\cdot {\bf v}_{\bf q})$ leads to \emph{multi-angle} effects \cite{Duan:2006an}, i.e., neutrinos moving on different trajectories experience different potentials.

The last term on right-hand-side in Eq.\,(\ref{eq:eom}) represents a collisional term acting on neutrino flavor evolution if they are still undergoing collisions
with matter or amongst themselves. Collisions occur at a rate proportional to $G_F^2$. 
In the context of both MSW and collective flavor conversions, the collisional term is expected to be negligible, as the conversions occur far from the neutrinosphere, where
neutrinos are free-streaming. However, the situation is less clear for fast conversions. A back-of-the-envelope calculation, using a nucleon density $n_B = \rho_{\rm nuc}/m_N \approx 1.8 \times 10^{38}$~cm$^{-3}$ and the neutrino-nucleon scattering cross-section
$\sigma \sim G_F^2 E^2 \sim 10^{-42}$~cm$^{-2}$ for $E_\nu \sim 10$~MeV, suggests that the scattering rate is $\Gamma= \sigma n_B \sim 10^7$~s$^{-1}$.
 We will find fast conversions can occur with a larger rate $\sim 10^8$~s$^{-1}$
 and therefore neglect the collisional effects as a first approximation.  We leave a dedicated investigation of this to a future work.

Even after neglecting the collisions, a  self-consistent solution of the flavor evolution requires solving the complete space-time-dependent problem described by Eq.\,(\ref{eq:eom}). First attempts at solution, by Fourier transforming Eq.\,(\ref{eq:eom}) along some of the space or time directions, have been recently presented in~\cite{Mangano:2014zda,Mirizzi:2015fva,Duan:2014gfa,Chakraborty:2015tfa,Dasgupta:2015iia,Abbar:2015fwa,Capozzi:2016oyk}. However, with the tools available at present, solving the full seven-dimensional problem remains a formidable challenge. 
 
\begin{figure}[!t]
\begin{centering}
\includegraphics[height=6.0cm]{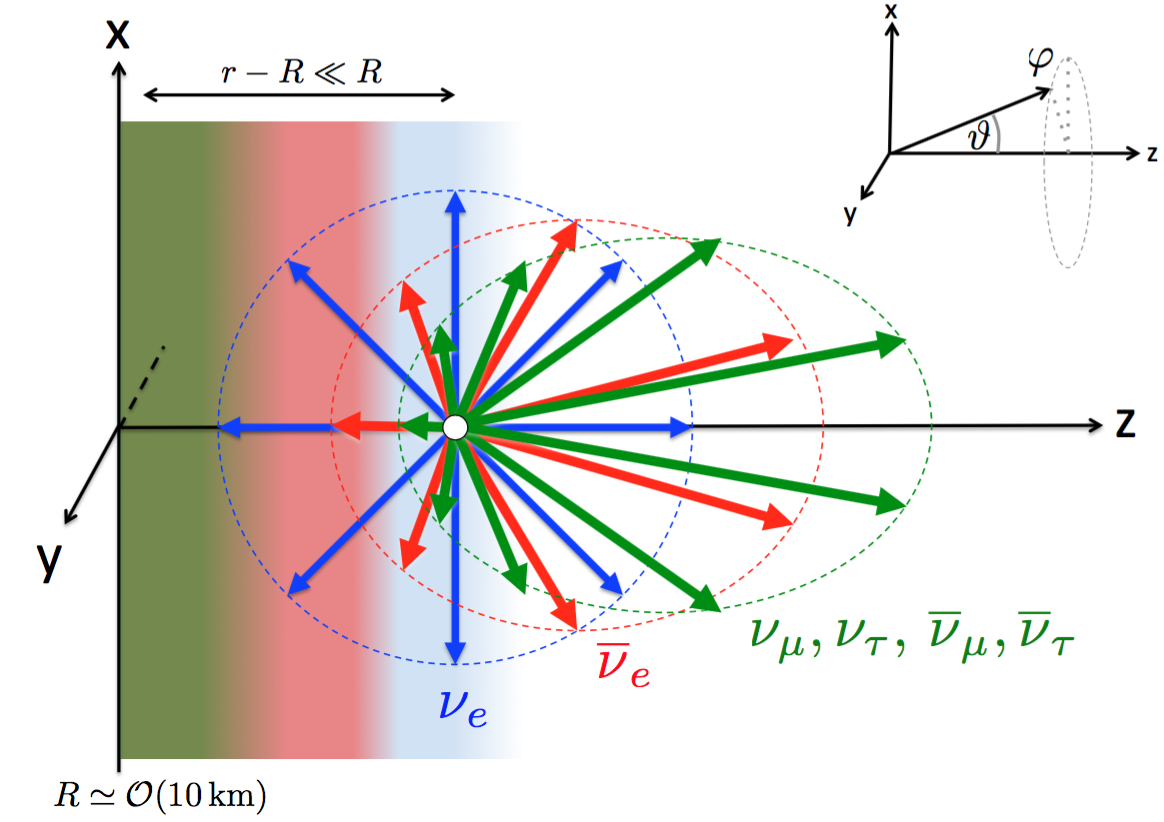}
\end{centering}
\caption{Schematic geometry of the model and flavor-dependent zenith-angle distributions of neutrino fluxes. The 3 ellipses are schematic polar plots of the normalized angular distributions of the $\nu_e$ (blue), $\bar{\nu}_e$ (red), and $\nu_x$ (green) fluxes at the point where the arrows originate.}
\label{fig:1}
\end{figure}

Interestingly, a major simplification suggests itself if one is interested in studying flavor conversions only at small distances from the SN core. Most of the neutrinos are emitted around a radius ${\cal{O}}(10)\,{\rm km}$ from the center of the SN. For phenomena that take place very close to this emission region, the curvature of the neutrinosphere is not relevant. We therefore model the source region as a diffuse flat infinite plane, as shown in Fig.\,\ref{fig:1}.

The neutrinos are conveniently labelled by $\omega,\,v_z,\,{\rm and}\,\varphi$, that define the Cartesian components of the momenta
\begin{equation}
{\bf p}=\Big(E\sqrt{1-v_z^2}\cos\varphi,E\sqrt{1-v_z^2}\sin\varphi,Ev_z\Big)\,\ ,
\end{equation}
where $v_z=\cos\vartheta$ is the component of the neutrino velocity along the $z$-axis, and $\vartheta$ and $\varphi$ the zenith and azimuthal angles, respectively. Note that $v_z$ can take negative values, i.e., the zenith angle $\vartheta$ can take values between $0$ and $\pi$, not merely up to $\pi/2$ as usually taken in the ``bulb'' model, representing neutrinos with trajectories that range from radially outward to radially inward into the star. 

The state of the neutrino population can then be represented as
\begin{equation}
\varrho_{\omega,v_z,\varphi}=\frac{1}{2}{\rm Tr}(\varrho_{\omega,v_z,\varphi})\,{\mathbb I}+\Phi_\nu\frac{g_{\omega,v_z,\varphi}}{2}\begin{pmatrix}
                                                \,s_{\omega,v_z,\varphi} && S_{\omega,v_z,\varphi}\\
                                                S_{\omega,v_z,\varphi}^* && -s_{\omega,v_z,\varphi}
                                               \end{pmatrix}\,,
\end{equation}
where $-1<s_{\omega,v_z,\varphi}<1$ and $s_{\omega,v_z,\varphi}^2+|S_{\omega,v_z,\varphi}|^2=1$. Neutrinos are produced as flavor eigenstates and no flavor mixing occurs as long as $S_{\omega,v_z,\varphi}=0$.  What is relevant for oscillations is the difference of the differential flux distributions for the two flavors, $g_{\omega,v_z,\varphi}\propto d\phi_{\nu_e}/d\Gamma-d\phi_{\nu_x}/d\Gamma$ for neutrinos and $\propto d\phi_{\nu_x}/d\Gamma-d\phi_{\bar\nu_e}/d\Gamma$ for antineutrinos represented by negative $\omega$, where we have assumed identical angular distributions
for $\nu_x$ and ${\bar\nu}_x$. This encodes the distribution of neutrino fluxes in energy, zenith angle, and azimuthal angle, i.e., over $d\Gamma=d\omega\,dv_z\,d\varphi/(2\pi)$, and $\Phi_\nu$ is the normalization of $g_{\omega,v_z,\varphi}$, determined by the condition
\begin{align}
\Phi_\nu\int_{-\infty}^{0}d\Gamma\,g_{\omega,v_z,\varphi}=-(\Phi_{\bar{\nu}_e}-\Phi_{{\nu}_x})\,,
\end{align}
where $\Phi_{\bar\nu_{e,x}}$ are the flavor-dependent total number fluxes averaged over the sphere of radius $\bf x$. This specific normalization is immaterial and all physical quantities depend only on the product $\Phi_\nu g_{\omega,v_z,\varphi}$. The neutrino-neutrino interaction potential is then given by $\mu=\sqrt{2}G_F\Phi_\nu$. 
The expected nature of the flavor-dependent neutrino distributions is sketched in Fig.\,\ref{fig:1}. We note that these differential fluxes $d\phi_{\nu_\alpha}/d\Gamma$ are predicted in some of the detailed SN simulations, and can be used as initial conditions for subsequent flavor evolution.

\section{Stability Analysis and Numerical Solutions for Schematic Neutrino Angular Distributions}
\label{sec:3}

While the self-induced flavor evolution represented by  Eq.\,(\ref{eq:eom}) is a non-linear phenomenon, the onset of these conversions can be examined through a standard \emph{stability analysis} of the linearized EoMs. The main idea behind this analysis is to determine if the neutrino flavor composition is stable, or does it have instabilities that grow exponentially with evolution in time or space? This technique, developed in~\cite{Banerjee:2011fj} (see also ref.~\cite{Sawyer:2008zs} for a related approach), reduces the nonlinear problem to a linear eigenvalue equation that involves the neutrino density, energy spectrum, angular distribution, matter density, etc. 

To linear order in $S_{\omega,v_z,\varphi}$, we have the EoMs
\begin{widetext}
\begin{eqnarray}
  i(\partial_t+v_z\partial_z+\vec{v}_T\cdot\partial_{T})S_{\w,v_z,\varphi}&=&\left[\w+\lam+\mu\int d\Gamma'\left(1-v_zv_z'-\vec{v}_T.\vec{v}_T'\right)g_{\w',v_z',\varphi'} \right]S_{\w,v_z,\varphi} \nonumber\\
                                               &&- \mu\int d\Gamma'\left(1-v_zv_z'-\vec{v}_T.\vec{v}_T'\right)g_{\w',v_z',\varphi'}\,S_{\w',v_z',\varphi'}\,,
\label{eq:stabeom}
\end{eqnarray}
\end{widetext}
where $\vec{v}_T$ is the velocity vector of the neutrino projected on the $x$-$y$--plane. For now, we assume translation invariance of the solutions along the transverse directions and drop the term involving the $\partial_T$; we will comment on this issue at the end this section. 

Our discussion pertains to a situation where $\mu\gg\Delta m^2/(2E)$ for all relevant neutrino energies $E$. We can then ignore the vacuum term, and integrate Eq.\,(\ref{eq:stabeom}) under $\int d\omega\,g_{\omega,v_z,\varphi}$ to find that the evolution of ${\tilde S}_{v_z,\varphi}\equiv\int d\omega\,g_{\omega,v_z,\varphi}\,S_{\w,v_z,\varphi}$ depends on spectrum $g_{\omega,v_z,\varphi}$ only through $\tilde{g}_{v_z,\varphi}\equiv\int d\omega\,g_{\omega,v_z,\varphi}$, i.e., the difference of the flux-weighted angular spectra of the neutrinos and antineutrinos. Explicitly, when $g_{\w,v_x,\varphi}\propto (d\phi_{\nu_e}/d\Gamma-d\phi_{\nu_x}/d\Gamma)_{\omega>0} + (d\phi_{\bar\nu_e}/d\Gamma-d\phi_{\bar\nu_x}/d\Gamma)_{\omega<0}$ is integrated over $\omega$, the $\nu_x$ and $\bar\nu_x$ dependent terms cancel each other. As a result, the $\nu_x$ and $\bar\nu_x$ distributions do not enter the EoMs as long as they are equal\footnote{We are grateful to Georg Raffelt for clarifying this issue to us.}. At this point, one could integrate out $\omega$ and study the stability of $\tilde{S}_{v_z,\varphi}$. However, we will study the stability of $S_{\w,v_z,\varphi}$, keeping our equations general and explicitly retaining $\omega$, setting it to zero only at the end.

\begin{figure}[!t]
\begin{centering}
\includegraphics[height=2.5cm]{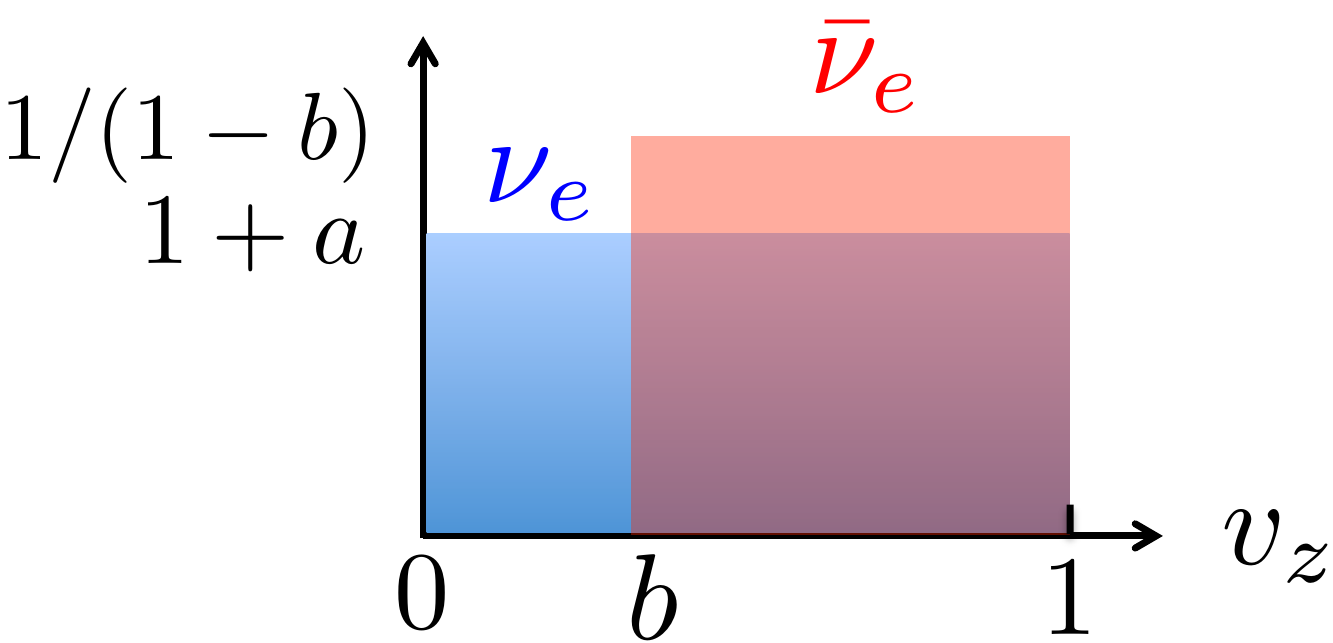}\hspace{1.0cm}\includegraphics[height=2.5cm]{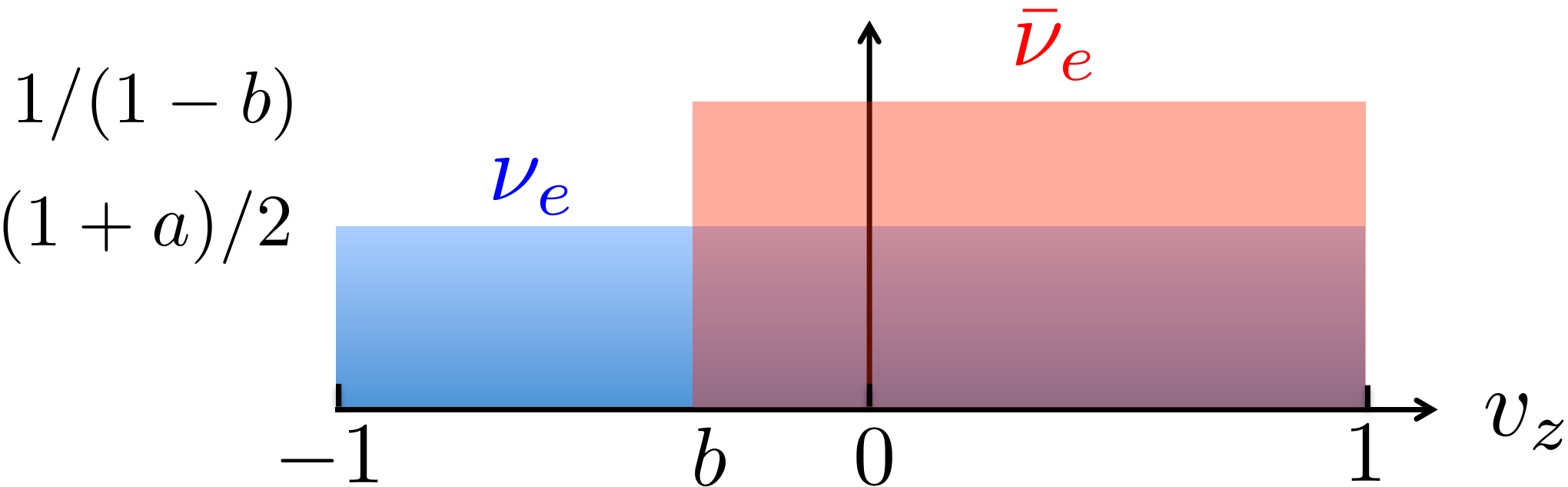}
\end{centering}
\caption{Sketches of the schematic zenith angle distributions of $\nu_e$ (blue) and $\bar\nu_e$ (red), used for the calculations in this section. The left panel shows a spectrum that corresponds to Eq.\,(\ref{eq:spectrum}) with no ingoing $\nu_e$ or $\bar\nu_e$, while the right panel shows a spectrum with ingoing $\nu_e$ and $\bar\nu_e$ as in Eq.\,(\ref{eq:spectrum2}). The $\nu_e$ and $\bar{\nu}_e$ have a flux ratio $1+a$, i.e., more $\nu_e$ than $\bar\nu_e$ when $a>0$, and the $\bar\nu_e$ have a more forward-peaked distribution, controlled by the parameter $b$ which we always choose to be larger than the ${\rm min}(v_z)$ for $\nu_e$.}
\label{fig:2}
\end{figure}

In this section, we consider only $\nu_e$ and $\bar\nu_e$ with the spectrum
\begin{equation}
g_{\omega,v_z,\varphi}=\frac{1}{2\pi}\left[(1+a)f_{\nu_e}(\omega)\Theta(v_z)\Theta(1-v_z)-\frac{1}{(1-b)}f_{\bar\nu_e}(\omega)\Theta(v_z-b)\Theta(1-v_z)\right]\,\ ,
\label{eq:spectrum}
\end{equation}
which, once integrated over their normalized $\omega$-distributions $f_{\nu_e}(\omega)$ and $f_{\bar\nu_e}(\omega)$, encodes the difference of the flux-weighted zenith-angle distributions of $\nu_e$ and $\bar\nu_e$. These distributions are individually shown in Fig.\,\ref{fig:2} (left panel).
The factor $1+a$ encodes the ratio of the total neutrino to antineutrino flux, which we expect to be larger than 1. 
Note that the spectrum is independent of $\varphi$. However, we shall find  that this azimuthal symmetry often gets spontaneously broken. Most importantly, however, the zenith angle distributions for the neutrinos and antineutrinos are not the same. While neutrinos are emitted over the entire forward hemisphere ($0\leq v_z\leq 1$), the antineutrinos are contained in a narrower forward cone $b\leq v_z\leq 1$, with $b>0$. As long as $1/(1-b)>1+a$, there is a \emph{crossing} of the two flux-weighted angular spectra. This kind of a ``non-trivial'' flavor-dependent angular distribution {is believed to be} be crucial for fast conversion. 

We will also investigate the impact of extending the range of $v_z$ to negative values, i.e., ($-1\leq v_z\leq 1$) for neutrinos and with $b>-1$, as shown in the right panel of Fig.\,\ref{fig:2}, to understand the role of inward going neutrinos and antineutrinos. However, we limit our focus to physically motivated spectra such that $\nu_e$ have larger fluxes and wider distributions in the zenith angle, compared to $\bar\nu_e$.

\subsection{Stationary Solutions with Evolution in Space}
We begin by looking for a steady state or stationary solution, i.e., the density matrices do not change with time. In that case it is appropriate to drop the time-derivative in Eq.\,(\ref{eq:stabeom}), and the eigenvalue equation for $S= Qe^{-i\Omega z}$ becomes
\begin{eqnarray}
  &&\left[\w+\lam+\mu\epsilon-(\Omega+\mu\epsilon_v)\,v_z - \mu\left(\epsilon_{c}\cos\varphi+
  \epsilon_{s}\sin\varphi\right)\sqrt{1-v_z^2}\right]Q=\nonumber\\ 
  &&\quad\quad\quad\quad\quad\mu\int { d\Gamma'}\left[1-v_zv_z'-\cos(\varphi-\varphi')\sqrt{(1-v_z^2)(1-v_z'^{\,2})}\right]g_{\w',v_z',\varphi'}\,Q'\,\ ,
   \end{eqnarray}
where we have defined the integrals over the spectrum,
\begin{eqnarray}
\epsilon        &=&\int d\Gamma\,g_{\w,v_z,\varphi}\,\ ,\\
 \epsilon_v      &=&\int d\Gamma\,v_z\,g_{\w, v_z,\varphi}\,\ , \\
\epsilon_{s(c)} &=&\int d\Gamma\,{\sin}{\varphi}\,(\cos{\varphi})\sqrt{1-v_z^2}~g_{\w,v_z,\varphi}\,\ ,
\end{eqnarray}
that encode the total, zenith, and azimuthal asymmetries, respectively. We make the ansatz that
\begin{equation}
Q=\frac{q_1+q_2\,v_z+q_3\,\cos\varphi\sqrt{1-v_z^2}+q_4\,\sin\varphi\sqrt{1-v_z^2}}{\w+\lam+\mu\epsilon-(\Omega+\mu\epsilon_v)\,v_z - \mu\left(\epsilon_{c}\cos{\varphi}+
  \epsilon_{s}\sin{\varphi}\right)\sqrt{1-v_z^2}}\,\ ,
\end{equation}
which gives us an eigenvalue equation,
\begin{equation}
\renewcommand*{\arraystretch}{1.6}
\left[ \begin{array}{c} q_1 \\ q_2 \\ q_3\\ q_4 \end{array} \right] = \begin{bmatrix}  I^{0,0}_{0,0} &  I^{0,0}_{1,0} &  I^{1,0}_{0,1} &  I^{0,1}_{0,1}\\ 
  -I^{0,0}_{1,0} &  -I^{0,0}_{2,0} &  -I^{1,0}_{1,1} &  -I^{0,1}_{1,1}\\
  -I^{1,0}_{0,1} &  -I^{1,0}_{1,1} &  -I^{2,0}_{0,2} &  -I^{1,1}_{0,2}\\ 
   -I^{0,1}_{0,1} & -I^{0,1}_{1,1} &  -I^{1,1}_{0,2} &  -I^{0,2}_{0,2} \end{bmatrix}\left[ \begin{array}{c} q_1 \\ q_2 \\ q_3\\ q_4 \end{array} \right]\,\ ,
\renewcommand*{\arraystretch}{1.0}
\label{eq:stabequ}
\end{equation}
in terms of a family of integrals
\begin{equation}
 I^{\alpha,\beta}_{m,n}=\mu\int d\Gamma\left[\frac{\cos^{\,\alpha}\varphi\,\sin^{\,\beta}\varphi~v_z^m~{(1-v_z^2)}^{\,n/2}}
 {\w+\lam+\mu\epsilon-(\Omega+\mu\epsilon_v)\,v_z - \mu\left(\epsilon_{c}\cos\varphi+
  \epsilon_{s}\sin\varphi\right)\sqrt{1-v_z^2}}\right]\,g_{\w,v_z,\varphi}\,\ .
\label{eq:Iint}
\end{equation}
Note that the integrals are dimensionless, and functions of $\Omega$ and other parameters.

The main point here is that whether fast flavor conversions take place depends on whether Eq.\,(\ref{eq:stabequ}) has complex solutions for $\Omega$. The imaginary part of $\Omega$, that we denote as usual as $\kappa={\rm Im}(\Omega)$, leads to an exponential rise in $S_{\w,v_z,\varphi}\sim e^{\kappa r}$. If ${\rm Im}(\Omega)$ happens to be nonzero in the limit of vanishing $\omega$ and $\lambda$, it can only be proportional to $\mu$, which is the only remaining dimensionful scale, signalling an instability whose rate scales directly with $\mu$, in contrast to the usual bipolar instabilities that scale as $\sqrt{\omega\mu}$. 

If $g_{\w,v_z,\varphi}$ is independent of $\varphi$, as we have chosen in Eq.\,(\ref{eq:spectrum}), the $\varphi$ integrals decouple and we can drop the indices $\alpha,\beta$ in the integrals $I^{\alpha,\beta}_{m,n}$ (setting them to zero), and write Eq.\,(\ref{eq:stabequ}) as
\begin{equation}
\renewcommand*{\arraystretch}{1.2}
 \left[ \begin{array}{c} q_1 \\ q_2 \\ q_3\\ q_4 \end{array} \right] = \begin{bmatrix}  
 I_{0,0} &  I_{1,0} &  0 &  0\\ 
  -I_{1,0} &  -I_{2,0} & 0 & 0\\
  0 &  0 &  -I_{0,2}/2 &  0\\ 
   0 & 0 &  0 &  -I_{0,2}/2 \end{bmatrix}\left[ \begin{array}{c} q_1 \\ q_2 \\ q_3\\ q_4 \end{array} \right]\,\ .
\renewcommand*{\arraystretch}{1.0}
\end{equation}
The upper $2\times 2$ block, manifestly independent of $\varphi$, gives the azimuthally symmetric solution, whereas the diagonal lower block gives the azimuthal symmetry breaking solution. The eigenvalues for the azimuthally symmetric instabilities are given by
\begin{equation}
 \left(I_{0,0}-1\right)\left(I_{2,0}+1\right)-(I_{1,0})^2=0\,\ ,
 \label{eq:stab1}
\end{equation}
while for the azimuthally non-symmetric instabilities one has
\begin{equation}
 \left(\frac{I_{0,2}}{2}+1\right)=0 \,\ .
  \label{eq:stab2}
\end{equation}
These azimuthal symmetry breaking instabilities spontaneously can generate large $\varphi$-dependent variations in the flavor composition, even if one starts with almost perfectly symmetric initial condition. 

With a choice of the spectrum $g_{\w,v_z,\varphi}$ given in Eq.\,(\ref{eq:spectrum}), it is possible to perform the integrals $I_{m,n}$ analytically and one can write the eigenvalue equations explicitly. However, these are transcendental equations in $\Omega$ and one cannot obtain closed-form solution for $\Omega$ using them, in general. Thus we resort to solving Eqs.~(\ref{eq:stab1}) and (\ref{eq:stab2}) numerically using {\sc Mathematica}. 

\begin{figure}
\begin{centering}
\includegraphics[width=0.32\textwidth]{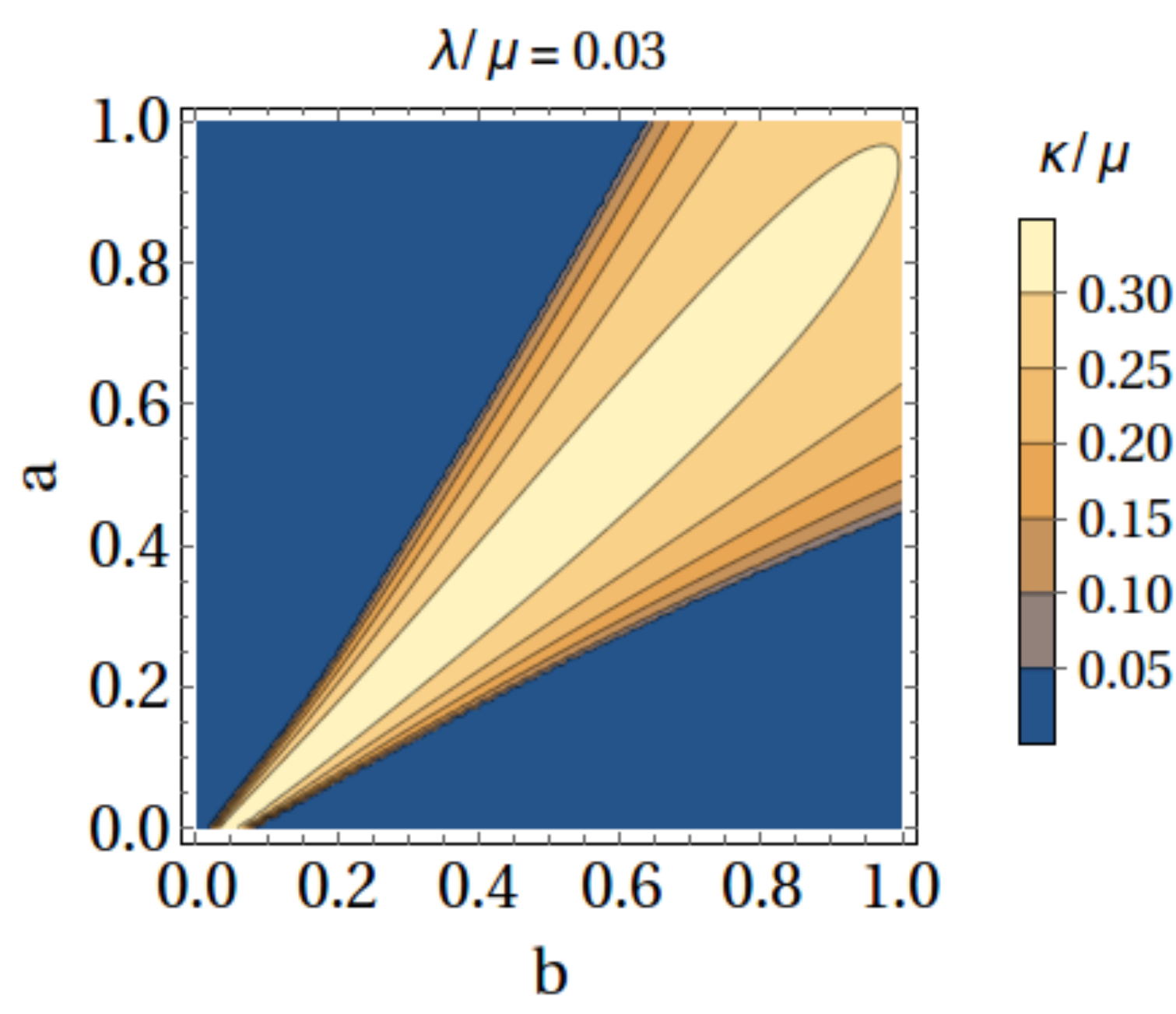}~\includegraphics[width=0.31\textwidth]{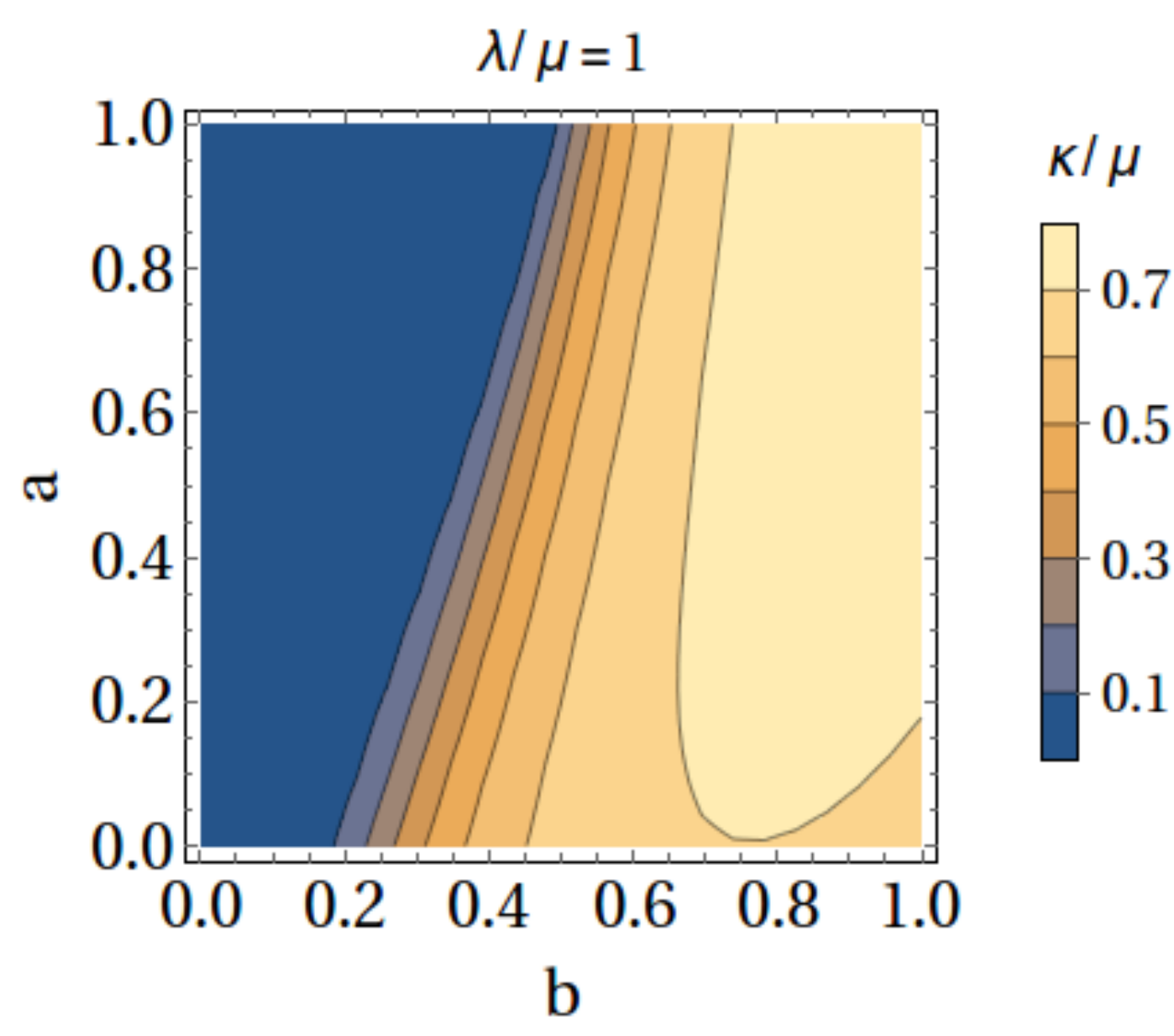}~\includegraphics[width=0.32\textwidth]{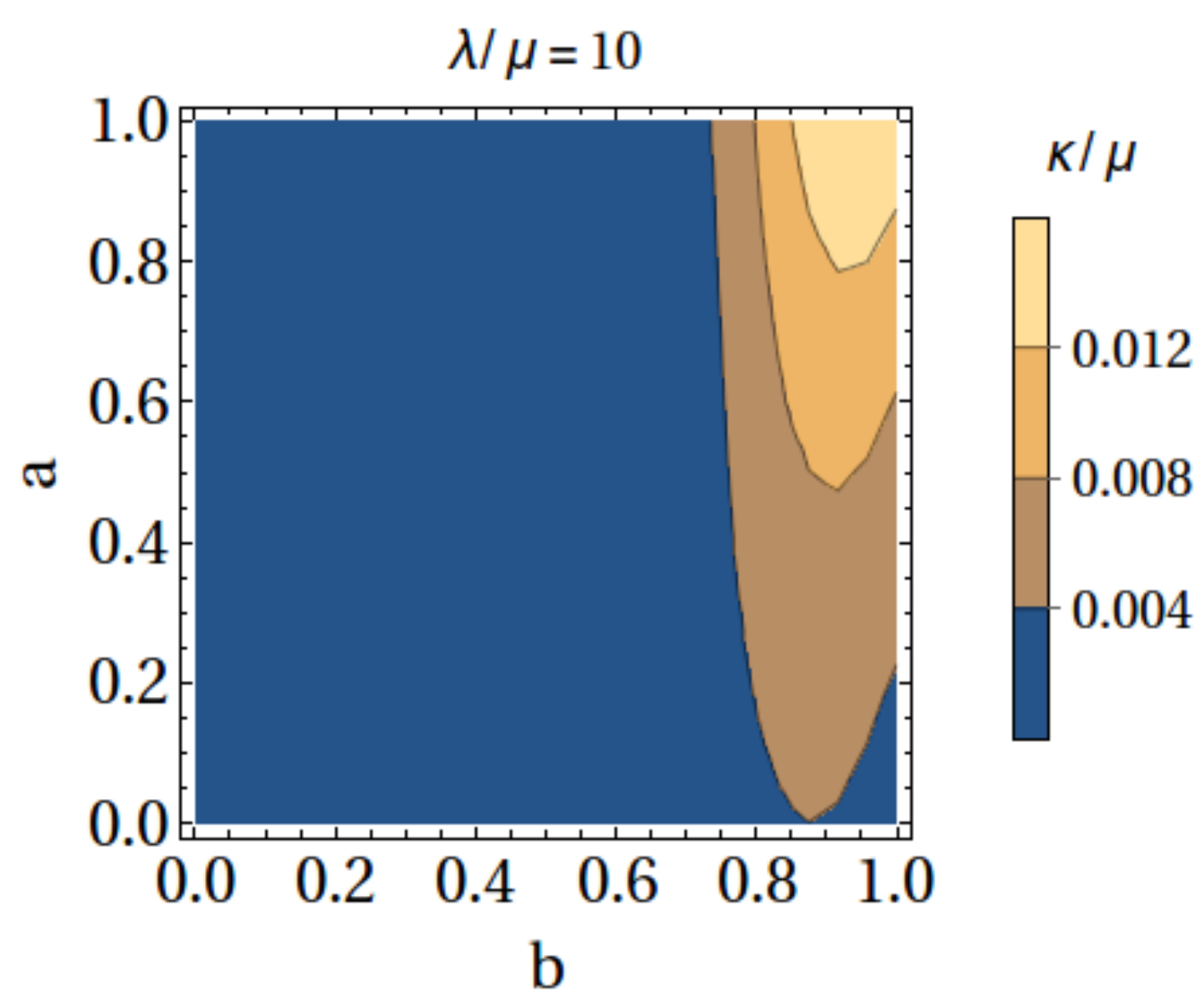}
\end{centering}
\caption{Instability rates for different values of $a$ and $b$, for three different values of $\lambda/\mu=$ 0.03, 1, and 10, respectively from left to right. Note that a large $a$ corresponds to large total flavor asymmetry $\epsilon$ and a zero $b$ corresponds to identical angular distributions for the two flavors, where the instabilities vanish. The instabilities are azimuthally symmetric, and we find no instabilities that break the azimuthal symmetry. For $\lambda=0$ we find no instabilities, azimuthally symmetric or not, that scale as $\mu$.}
\label{fig:3}
\end{figure}

In Fig.\,{\ref{fig:3}}, we show a contour plot of the imaginary part of $\Omega$ for different values of $a$ and $b$ which shows that fast conversions do not occur if $b=0$. The instabilities are azimuthally symmetric, i.e., solutions to Eq.\,(\ref{eq:stab1}) and we found no solutions to Eq.\,(\ref{eq:stab2}) with imaginary parts. We remind that the parameters $a$ and $b$, that define the spectrum $g_{\w,v_z,\varphi}$, are in fact closely related to the total and zenith angle asymmetries between the neutrino flavors,
\begin{equation}
\epsilon=a{\quad\rm and \quad}\epsilon_{v}=(a-b)/2\,\ ,
\end{equation}
For $\lambda=0$, we found no instabilities to either equation that had imaginary parts. These results are in qualitative agreement with the results obtained in ref.~\cite{Chakraborty:2016lct}%
\footnote{Repeating the calculations of ref.~\cite{Chakraborty:2016lct}, we found numerical agreement. However, we found the choice of range of $u$ to be confusing: It is allowed to take values beyond 1, as $u = 1 \pm b$ with $ -1\leq b \leq 1 $. Evidently, $u$ is defined therein as not exactly equal to $\sin^2\vartheta$ but merely proportional to it, including the radius-dependent factor of $r^2/R^2$. This factor was overlooked in their definitions of $\lambda$ and $\mu$, but it thankfully drops out of the ratio $\lambda/\mu$ and results remain unchanged.}.

The most important feature to be noticed here is that the are no instabilities if $b=0$, i.e., when the neutrino and antineutrino distributions are the same.
Therefore, a \emph{necessary} condition to have fast instabilities appears to be a \emph{crossing} between the angular spectra of $\nu_e$ and $\bar\nu_e$, as
shown in Fig.\,{\ref{fig:2}}.
 Another important feature to be noted is that instabilities exist only if $\lambda\sim\mu$, disappearing for both smaller and larger $\lambda$, thus raising some doubt if the instability rate does indeed scale as $\mu$, and not some finely tuned combination of $\lambda$ and $\mu$. One may wonder, if there is a cleaner demonstration that there are fast instabilities that are strictly proportional to $\mu$. We shall provide such an example in the next section.

\pagebreak[4]
\subsection{Homogeneous Solutions with Evolution in Time}
Now we turn our attention to the possibility that the flavor composition 
does not vary spatially, but undergoes rapid turnovers in time.
This is motivated by the results of the previous section, where we saw 
that unless $\lambda\sim\mu$, we did not find fast conversions. 
We will now show that even in such conditions, fast conversions  
may occur efficiently.

If we assume that the neutrino flavor composition is relatively 
homogeneous over the region of interest and only varies with time, we can drop the 
spatial derivatives in Eq.\,(\ref{eq:stabeom}), and write $S_{\w,v_z,\varphi}= Qe^{-i\Omega t}$. Thereafter, 
analogous to the previous section, one obtains the same eigenvalue equation for $Q$ as
in Eq.\,(\ref{eq:stabequ}), but with the integrals $I_{m,n}^{\alpha,\beta}$ replaced by a new 
family of integrals
\begin{equation}
  J^{\alpha,\beta}_{m,n}=\mu\int d\Gamma\left[\frac{\cos^{\,\alpha}\varphi\,\sin^{\,\beta}\varphi~v_z^m~{(1-v_z^2)}^{\,n/2}}
 {\w+\lam+\mu\epsilon-\Omega-\mu\epsilon_v\,v_z - \mu\left(\epsilon_{c}\cos\varphi+
  \epsilon_{s}\sin\varphi\right)\sqrt{1-v_z^2}}\right]\,g_{\w,v_z,\varphi}\,\ ,
\label{eq:Jint}
\end{equation}
which differ by the replacement 
$v_z\Omega\rightarrow\Omega$ in the denominator of the integrand of 
$I^{\alpha,\beta}_{m,n}$.

For a spectrum which is independent of $\varphi$, Eq.\,(\ref{eq:stabequ}) simplifies as before. The upper block gives the azimuthally symmetric solution whereas the lower block gives the azimuthal symmetry breaking solution. The eigenvalues for the azimuthally symmetric instabilities are given by
\begin{equation}
 \left(J_{0,0}-1\right)\left(J_{2,0}+1\right)-(J_{1,0})^2=0\,\ ,
\label{eq:J1}
\end{equation}
while for the azimuthally non-symmetric instabilities one has
\begin{equation}
 \left(\frac{J_{0,2}}{2}+1\right)=0 \,\ .
\label{eq:J2}
\end{equation}

\begin{figure}[t]
\begin{centering}
\includegraphics[width=0.45\textwidth]{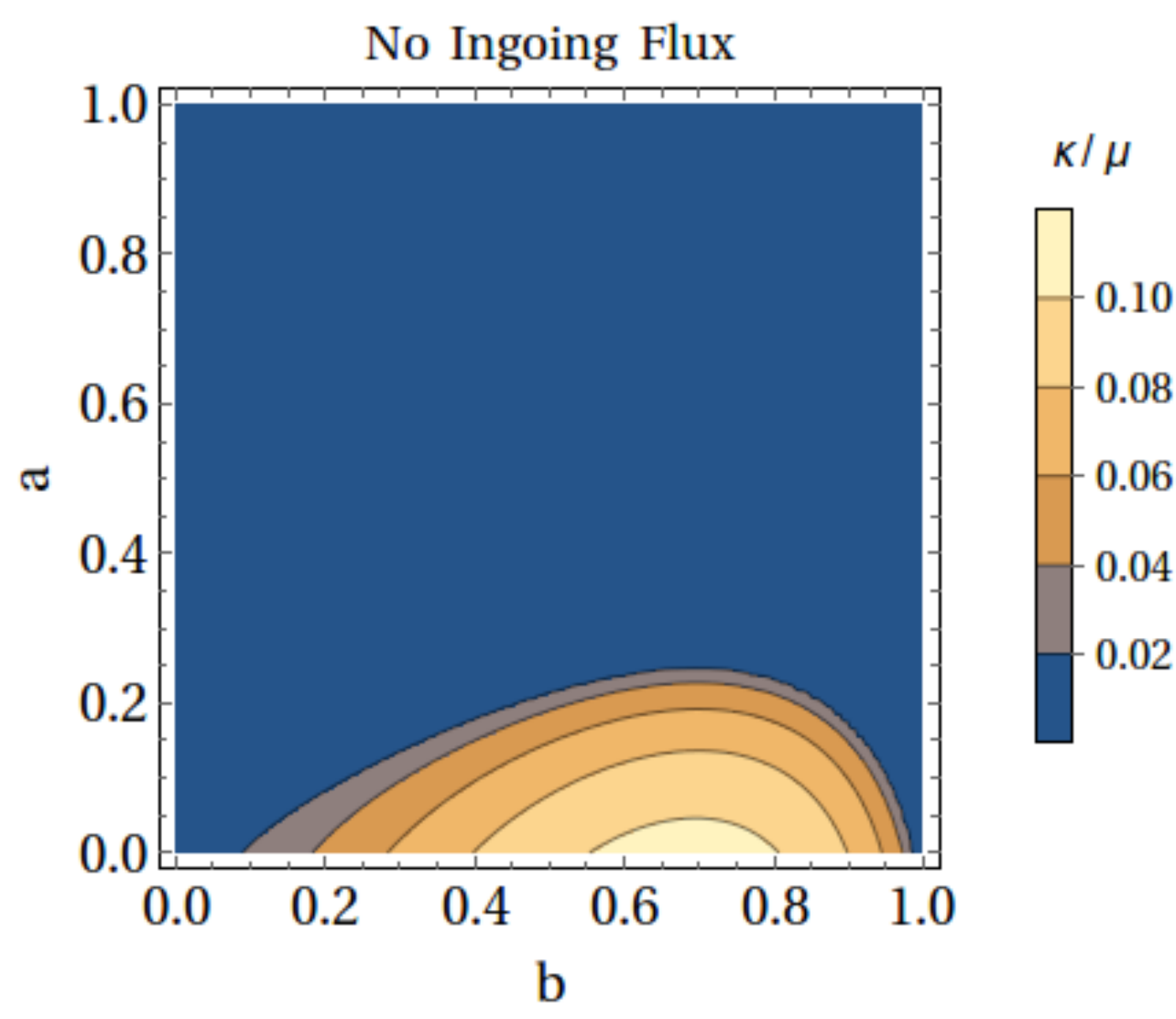}\hspace{1.0cm}\includegraphics[width=0.45\textwidth]{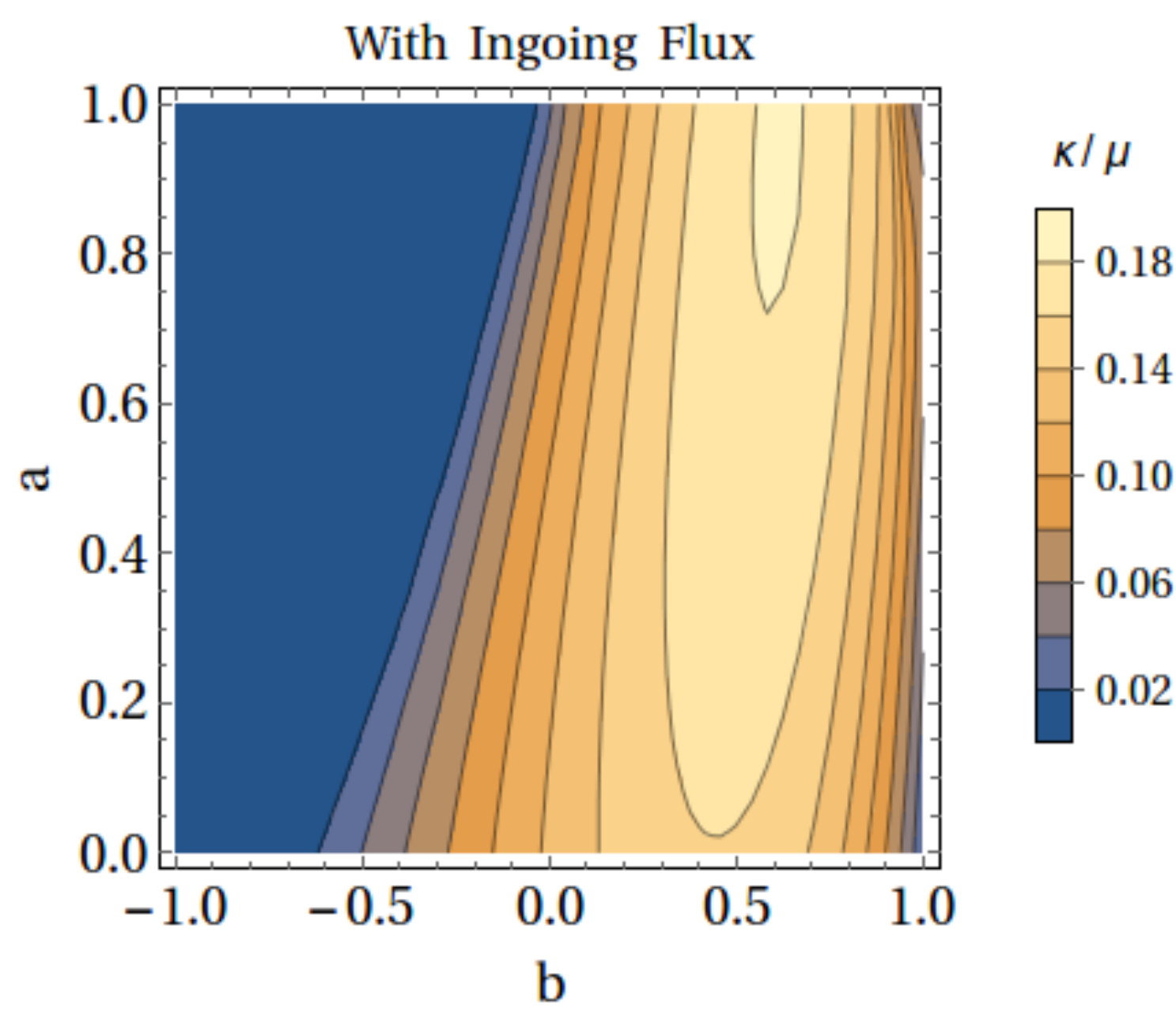}
\end{centering}
\caption{Instability rates for different values of $a$ and $b$, for evolution in time, without including inward going modes (left panel) and including inward going modes (right panel). These instabilities are azimuthally asymmetric, and we found no instabilities if the azimuthal symmetry were to be exact. There is no dependence on $\lambda$.}
\label{fig:4}
\end{figure}
\begin{figure}[h]
\begin{centering}
\includegraphics[width=0.45\textwidth]{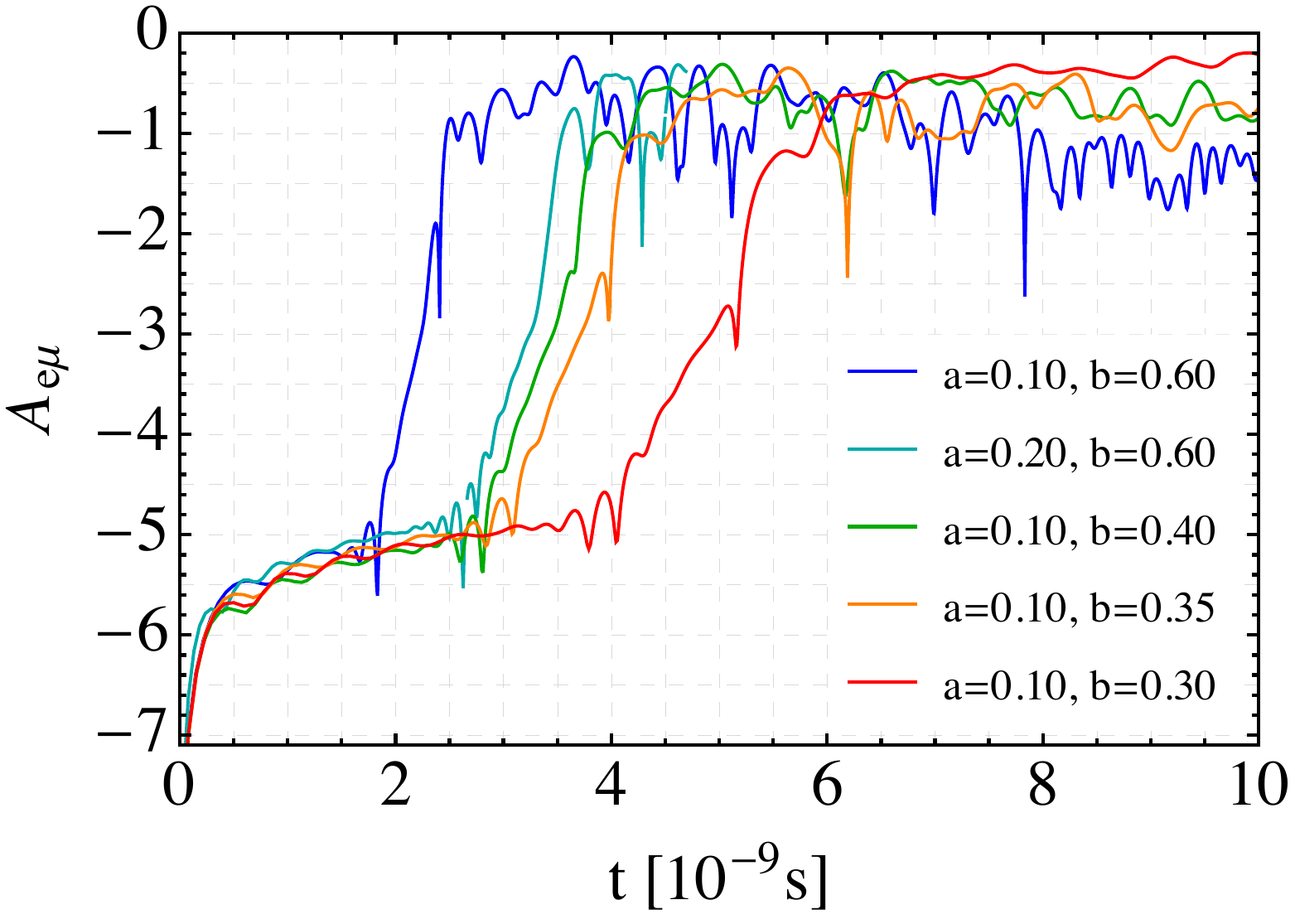}\hspace{1.0cm}\includegraphics[width=0.45\textwidth]{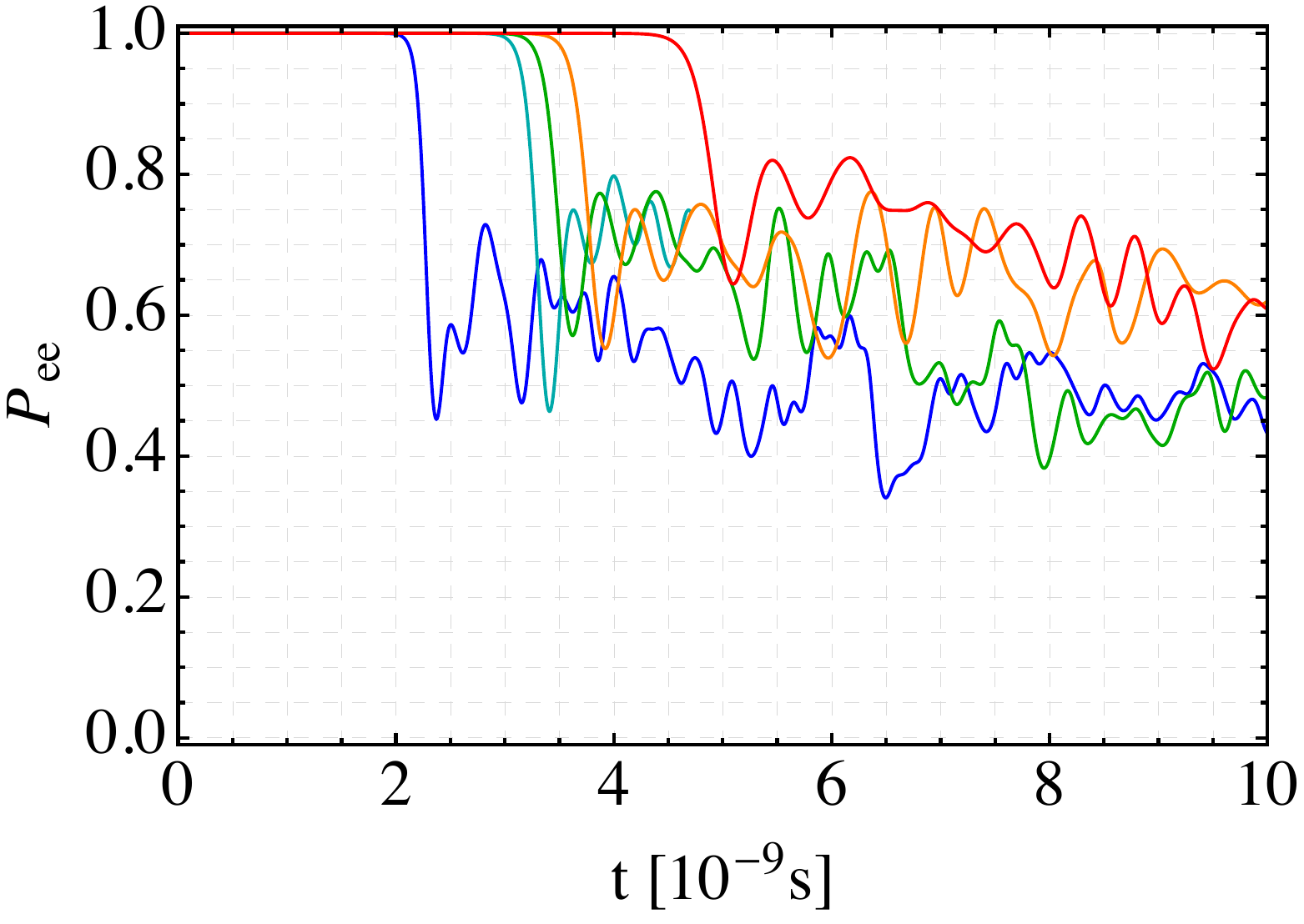}
\end{centering}
\caption{Growth of instability for evolution in time, as predicted by numerical solution of the nonlinear evolution of ${\bar\nu}_e$, for representative values of $a$ and $b$. The left panel shows quantity $A_{e\mu}={\log}_{10}|S|$ as a measure of the extent of flavor conversion. The right panel shows the   angle-integrated survival probabilities~$P_{ee}$. These instabilities are azimuthally asymmetric and independent of $\lambda$.}
\label{fig:5}
\end{figure}

An important fact here is that $\lambda$ does not affect the temporal stability in any crucial manner. If there is an unstable solution for $\lambda=0$, one will find an unstable solution with the same imaginary part for any other value of $\lambda$ by simply shifting the real part of $\Omega$, i.e., by shifting $\Omega\to\Omega+\lambda$, as is apparent from Eq.\,(\ref{eq:Jint}), which is the only manner in which $\lambda$ enters our equations.

In Fig.\,\ref{fig:4} (left panel) we show the instability rates for evolution in time, without including inward going modes, i.e., $g_{\omega,v_z,\varphi}$ is given by Eq.\,(\ref{eq:spectrum}). 
It is apparent that fast instabilities, which are azimuthal symmetry breaking solutions to Eq.\,(\ref{eq:J2}), exist only for small values of $a$ and large values of $b$. There is no dependence on $\lambda$, which can be absorbed into the real part of $\Omega$. 

On the right panel in Fig.\,\ref{fig:4}, we show the analogous results, but for the spectrum
\begin{equation}
g_{\omega,v_z,\varphi}=\frac{1}{2\pi}\left[\frac{1+a}{2}f_{\nu_e}(\omega)\Theta(1+v_z)\Theta(1-v_z)-\frac{1}{(1-b)}f_{\bar\nu_e}(\omega)\Theta(v_z-b)\Theta(1-v_z)\right]\,\ ,
\label{eq:spectrum2}
\end{equation}
where the $\nu_e$ are emitted isotropically along all zenith angles (see right panel of Fig.\,\ref{fig:2}). It is clear that, for the same value of $a$ and $b$, the presence of the backward travelling modes of $\nu_e$ greatly amplify the instabilities. What this means in practice is that closer to the neutrinosphere, the fast instability can be stronger due to the presence of these inward going neutrinos.

We have also numerically solved the fully nonlinear EoMs  for the spectrum corresponding to the left panel in Fig.\,\ref{fig:4} (no inward going modes). The EoMs were discretized in $v_z$ and $\varphi$, with 100 modes for $0 \leq v_z \leq 1$ and 10 modes in $\varphi$, and the $\nu$-$\nu$ interaction strength was taken to be $\mu= 4 \times 10^{5}$~km$^{-1}$. In Fig.\,\ref{fig:5} we show these numerically evaluated angle-integrated amplitudes of 
the flavor conversions for the ${\bar\nu}_e$,
\begin{equation}
A_{e \mu}(t) = \textrm{log}_{10}|S(t)|\,\ ,
\end{equation}
for some representative values of $a$ and $b$. The initial evolution, that asymptotes to a plateau at $A_{e\mu}\simeq10^{-5}$, is not the fast conversion predicted by linear stability analysis. However, the subsequent evolution, where $A_{e\mu}$ grows approximately linearly, is in excellent agreement with the corresponding growth rates shown in the left panel of Fig.\,\ref{fig:4}. We have also performed nonlinear calculations corresponding to the right panel of Fig.\,\ref{fig:4}, which show faster growth, but we do not show them here, as they are otherwise quite similar.
In all these cases we find a perfect agreement between the linear and non-linear calculation of $A_{e \mu}$. In particular, we find  exponentially growing run-away solutions within 
$t\sim {\mathcal O}(10^{-8})$\,s.

In the right panels we also show the corresponding angle-integrated survival probability given by $P_{e e}(t)$. As one can clearly see, fast conversions lead to approximate flavor equilibrium, i.e., $P_{ee}\simeq 1/2$. Depending on the details of these fast conversions, however, this equilibration may not necessarily be complete, e.g., if the instability growth rates are small.

\subsection{Evolution in both Space and Time}

If the density matrix evolves in both space and time, the formalism presented above is inadequate. A useful way of studying these solutions is to consider
\begin{equation}
S(\vec{x},t)=Q e^{-i(\vec{\Omega}_x\cdot\vec{x}+\Omega_t t)}\,\ ,
\end{equation}
and follow the approach in refs.~\cite{Dasgupta:2015iia,Capozzi:2016oyk} (see also~\cite{Chakraborty:2016yeg, Abbar:2015fwa}). One can Fourier transform all the $x,y,z,t$ dependencies, i.e., $Q e^{-i(\vec{\Omega}_x\cdot\vec{x}+\Omega_t t)}\to \sum_{\vec{k},p}Q_{\vec{k},p} e^{-i(\vec{k}\cdot\vec{x}+p t)}$, and study the stability of these Fourier modes. Physically, $\vec{k}$ and $p$ correspond to \emph{inhomogeneities} or \emph{pulsations} with wave-vectors $\vec{k}$ or frequency $p$, respectively. Here, we do not go into a detailed study along these lines, except to note a few key features:
\begin{itemize}
\item If we are looking at spatial evolution along $z$, the time-dependence in $S$ will appear as {pulsations} of frequency $p=\Omega_t$ that affects the linear stability, e.g., in Eq.\,(\ref{eq:stabequ}), through the replacement $\lambda\to\lambda-p$ in Eq.\,(\ref{eq:Iint}). 
 
\item  If on the other hand we wish to study evolution in time, the spatial oscillations of $S$ along $z$ can be Fourier decomposed into their constituent frequencies labelled by $k_z$, and one simply shifts
\begin{equation}
\mu\epsilon_v\to\mu\epsilon_v+k_z\,\ ,
\end{equation}
in Eq.\,(\ref{eq:Jint}). 
\item Fourier modes of the fluctuations along transverse directions $x$ and $y$ will lead to analogous shifts
\begin{equation}
\mu\epsilon_{c,s}\to\mu\epsilon_{c,s}+k_{x,y}\,\ ,
\end{equation}
in the denominator of Eq.\,(\ref{eq:Iint}) or Eq.\,(\ref{eq:Jint}).
\item The linear stability analysis can then proceed with these minor replacements, essentially amounting to these redefinitions of the asymmetry parameters. Practically, this may have interesting consequences leading to enhanced fast conversions, e.g., as was shown in refs.~\cite{Dasgupta:2015iia,Capozzi:2016oyk}, the pulsations can enhance instabilities by effectively removing the matter effect for the specific pulsating Fourier modes and the evolution of nearby modes via nonlinear coupling of modes. Similarly, inhomogeneity may dynamically mimic a larger zenith or azimuthal angle asymmetry and enhance fast conversions. A detailed exploration of these possibilities is left for a future study.

\end{itemize}


\section{Numerical solutions with SN Fluxes} 
\label{sec:4}
Numerical simulations of SN explosions predict flavor-dependent zenith-angle neutrino distributions, as expected on the physical grounds explained in Sec.\,\ref{sec:1}.
Inspired by SN simulations, a first attempt to characterize self-induced neutrino flavor conversions with  flavor-dependent $\nu$ distributions was performed in \cite{Mirizzi:2011tu,Mirizzi:2012wp}.
However, angular distributions for $\nu_e$ and ${\bar\nu}_e$ were assumed to be identical in these studies, and only a crossing between electron and non-electron neutrino angular spectra was considered. As a result, even though the flavor conversions were shown to be enhanced with respect to the case with trivial angular distributions, the possibility of fast conversions was missed. In light of the new insights gained with the toy-models considered in refs.~\cite{Sawyer:2005jk,Sawyer:2008zs,Sawyer:2015dsa,Chakraborty:2016lct} and in Sec.\,\ref{sec:3}, we now consider the possibility of fast conversions by introducing realistic SN angular distributions with different angular distributions for $\nu_e$ and $\bar\nu_e$.

\begin{figure}[t]
\begin{centering}
\includegraphics[width=0.45\textwidth]{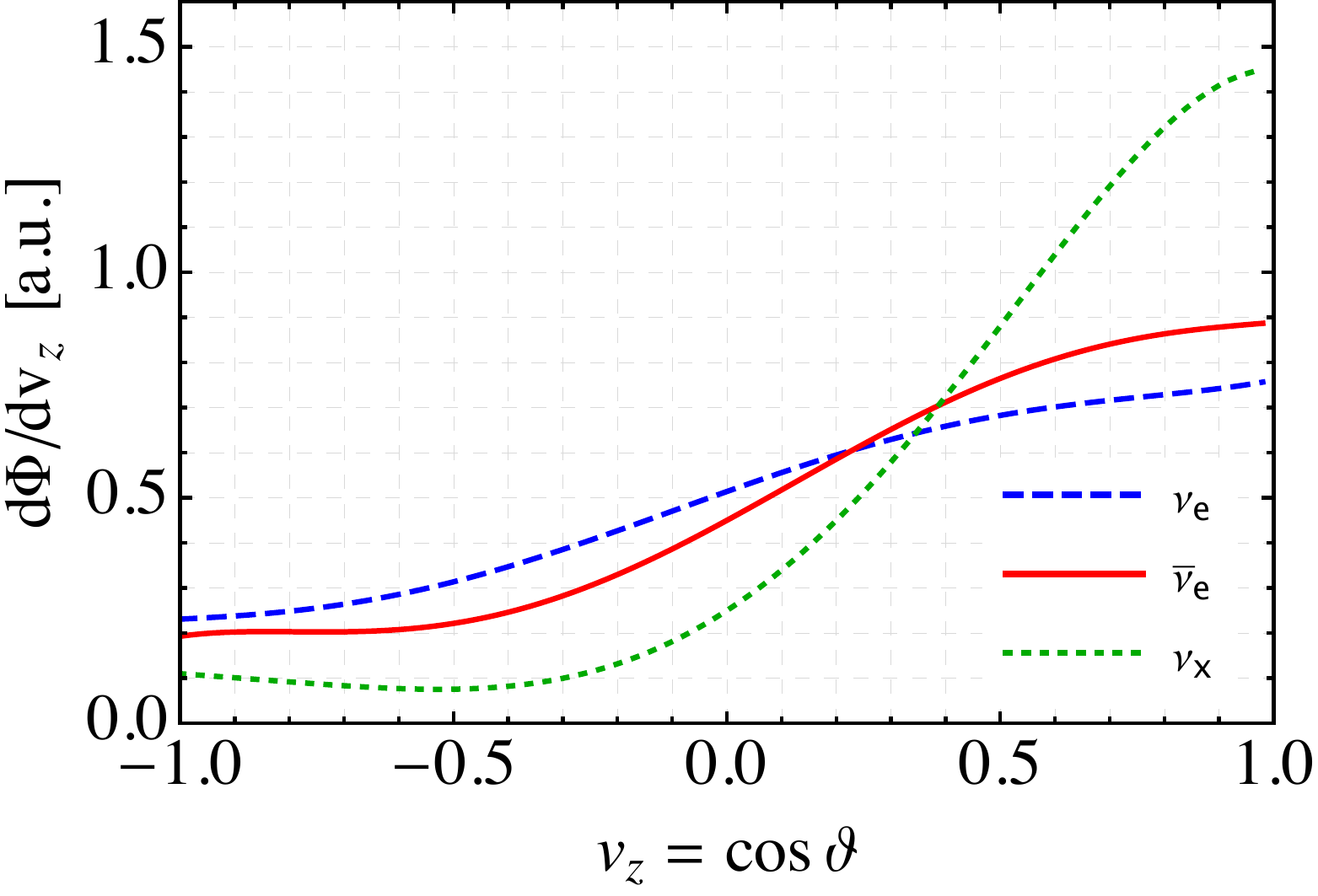}\hspace{1.0cm}
\includegraphics[width=0.45\textwidth]{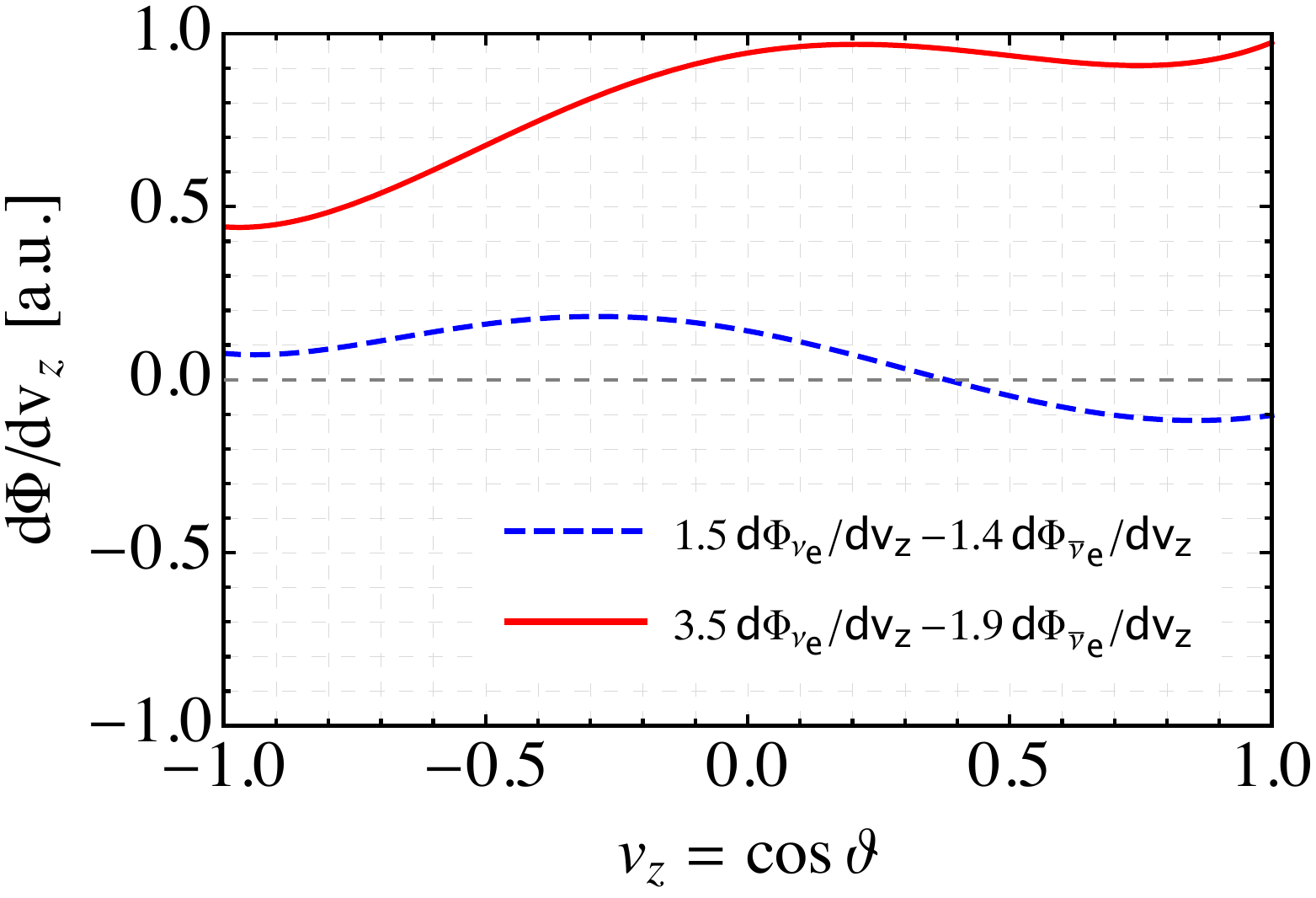}
\end{centering}
\caption{Left panel: Normalized flavor-dependent zenith angle distributions of SN neutrinos from a one-dimensional SN model from the Garching group for  a 25\,$M_{\odot}$ progenitor at post-bounce time $t=0.325$\,s and $r=25$\,km~\cite{garching}. Right panel: Difference of flux-weighted angular spectra of $\nu_e$ and $\bar\nu_e$, for two choices of flux ratios  corresponding to small asymmetry (dashed line) and large asymmetry (solid line), respectively. Note that $\nu_x$ and $\bar\nu_x$ fluxes are equal and thus drop out.}
\label{fig:6}
\end{figure}
 
At this stage we are not interested in an extensive phenomenological study, and we confine ourselves to single representative example. We take the normalized angular distributions for different $\nu$ species from a one-dimensional SN model for  a 25 $M_{\odot}$ progenitor at post-bounce time $t=0.325$~s and $r=25$~km, simulated by the Garching group~\cite{garching}, as shown in Fig.\,\ref{fig:6} (left panel). One notices that while the $\nu_x$ distributions are mostly forward-peaked, $\nu_e$ and $\bar\nu_e$ have a significant fraction of backward going neutrinos.

\begin{figure}[t]
\begin{centering}
\includegraphics[width=0.45\textwidth]{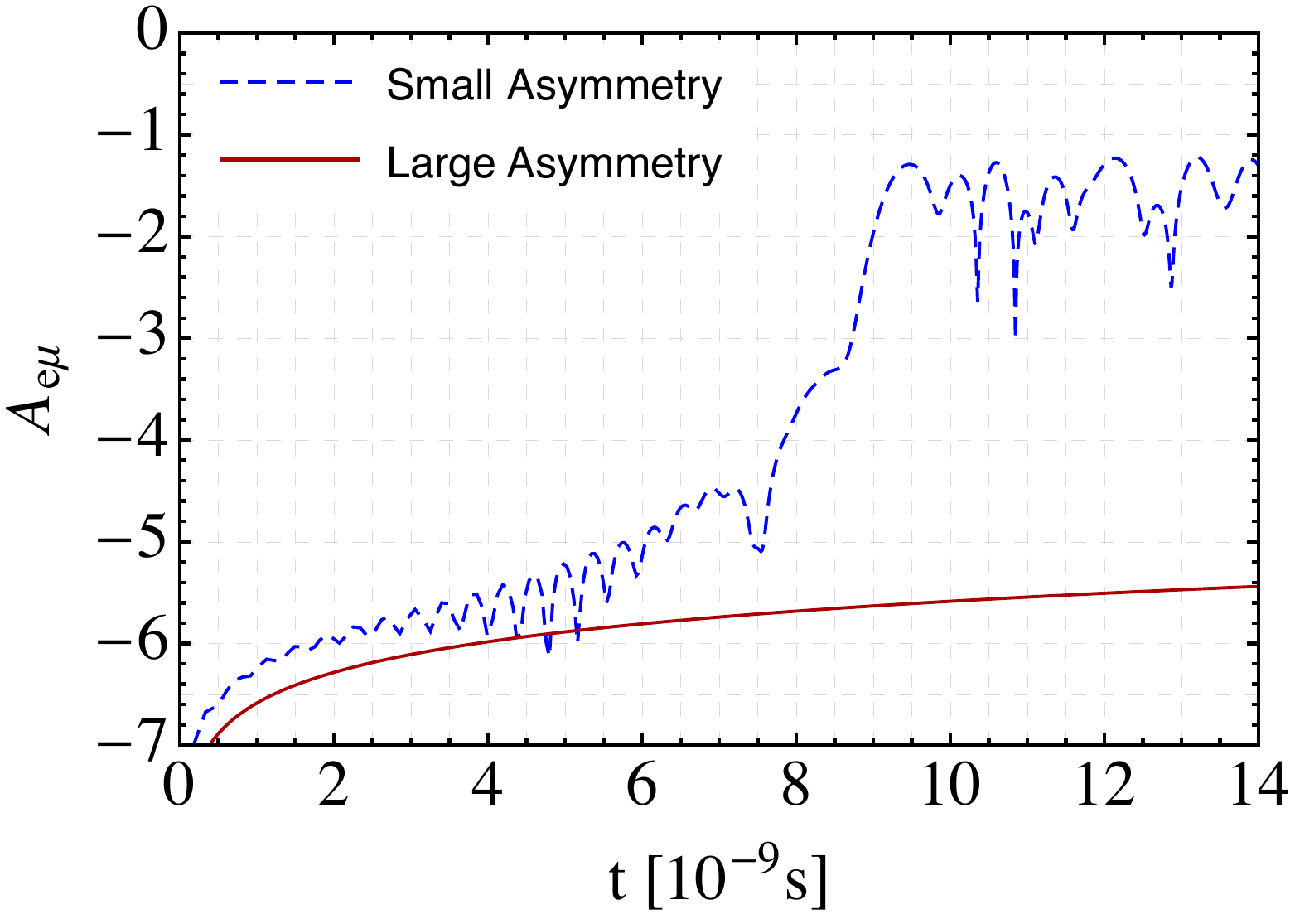}\hspace{1.0cm}\includegraphics[width=0.45\textwidth]{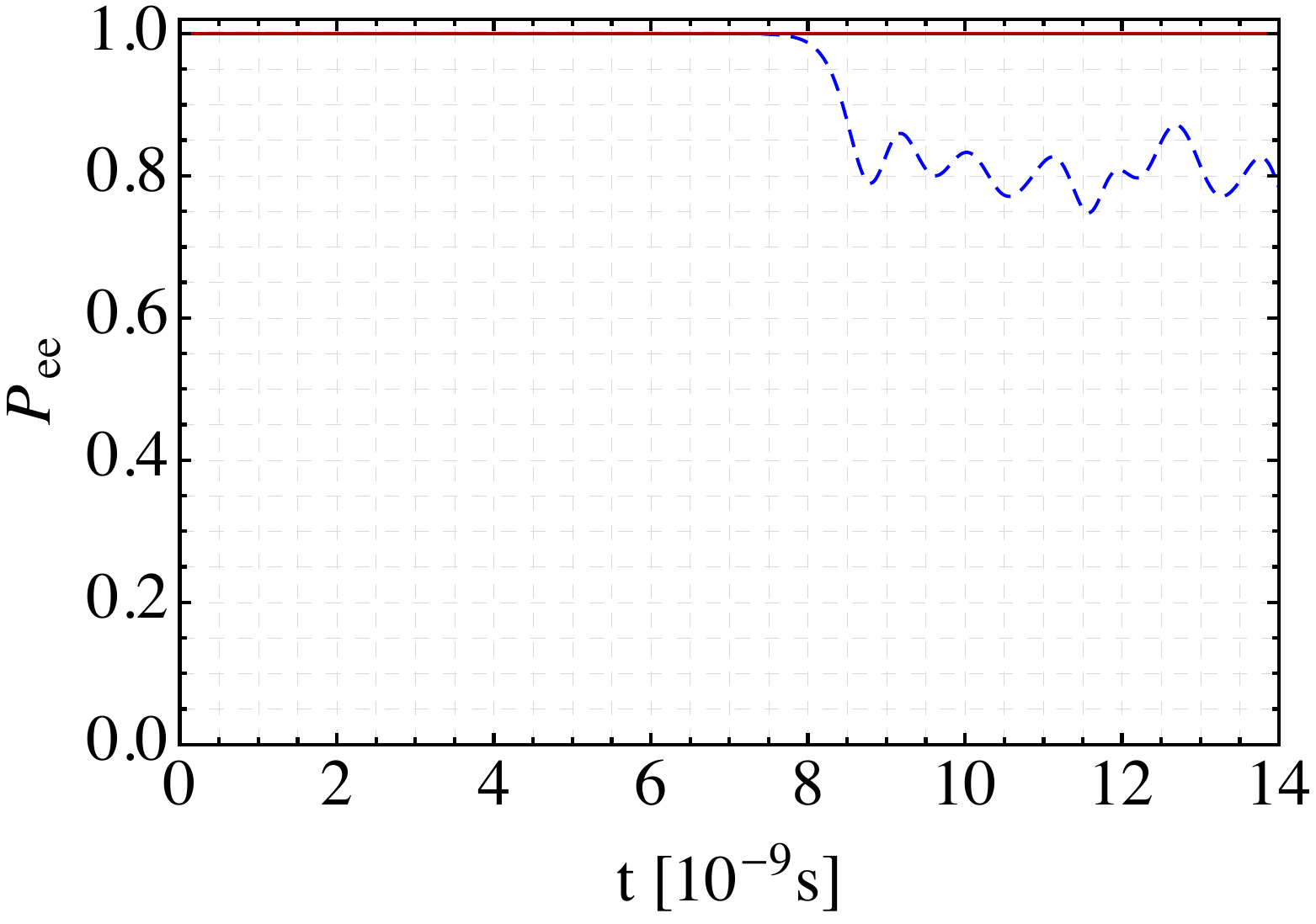}
\end{centering}
\caption{Growth of fast instabilities for realistic SN neutrino angular distributions considered in this paper. The dashed and continuous curves correspond to the flux parameters with small and large lepton asymmetries, as may be expected due to LESA. While large asymmetries suppress the fast conversion, for smaller asymmetries there is $\simeq20\%$ flavor conversion within a few nanoseconds. The growth of off-diagonal components is shown on the left panel, while the right panel shows the  angle-integrated survival probability for ${\bar\nu}_e$.}
\label{fig:7}
\end{figure}

In this specific simulation, one finds a strongly hierarchical flavor ratio $\Phi_{\nu_e}:\Phi_{\bar\nu_e}:\Phi_{\nu_x}=3.5:1.9:1$.
For such a strong $\nu_e$ to $\bar\nu_e$ asymmetry there is no crossing between the zenith-angle spectra of $\nu_e$ and $\bar\nu_e$ (see right panel in Fig.\,\ref{fig:6})  and we do not expect to find any fast conversion.
However, it has been recently discovered that multidimensional SN simulations exhibit a phenomenon called lepton-emission self-sustained asymmetry (LESA)~\cite{Tamborra:2014aua}, i.e., the lepton asymmetries of the neutrino fluxes have strong variance over various directions and this roughly hemispherical asymmetry appears to be self-stabilized. In particular, along some directions flavor asymmetries among the different species can be much milder than in the corresponding 1D simulations (see \cite{Chakraborty:2014lsa}).

Currently, the $\nu$ angular distributions for the multi-dimensional SN models showing the LESA effects are not available in a readily usable form, but they are expected to be similar to that in 1D simulations~\cite{jankaprivate}. Therefore, we use the 1D distributions shown in Fig.\,\ref{fig:6}, and simply change the relative weights of $\nu_e$ and $\bar\nu_e$ fluxes within the range predicted by models exhibiting LESA. In  particular we take two cases, one with a large asymmetry $\Phi_{\nu_e}:\Phi_{\bar\nu_e}:\Phi_{\nu_x}=3.5:1.9:1$, and another with a small asymmetry $\Phi_{\nu_e}:\Phi_{\bar\nu_e}:\Phi_{\nu_x}=1.5:1.4:1$, roughly corresponding to the direction with lowest asymmetry. 
In this latter case, one finds a crossing  between the zenith-angle spectra of $\nu_e$ and $\bar\nu_e$, as shown in the right panel of Fig.\,\ref{fig:6}. Therefore, fast conversions are expected here.
 
In Fig.\,\ref{fig:7} (left panel) we show the amplitude of flavor evolution $A_{e\mu}$ for ${\bar\nu}_e$, for these two choices of flux ratios. Note that these calculations include the inward going neutrino modes as shown in Fig.\,\ref{fig:6}. The right panel of Fig.\,\ref{fig:7} shows the corresponding  angle-integrated  survival probabilities $P_{e e}(t)$. One can see that while no fast conversion takes place for the large flux asymmetry, if the flux asymmetry is not very large fast conversion leads to almost ${\cal{O}}(1)$ flavor conversion within a few nanoseconds, in a range of $r-R\sim {\mathcal O}$(1)\,m from the boundary leading to approximate flavor equilibrium.

\section{Discussion and Outlook}
\label{sec:5}
In this paper we investigated the possibility of flavor conversions of SN neutrinos, that occur with 
a fast rate $\sim \mu$, driven by interactions of neutrinos emitted with non-trivial angular distributions. This was suggested in a series of papers~\cite{Sawyer:2005jk,Sawyer:2008zs,Sawyer:2015dsa,Chakraborty:2016lct}, that have stimulated our work. We have attempted to clarify the intriguing flavor dynamics by carefully delineating spatial and temporal flavor evolution, and considering physically well-motivated neutrino fluxes and emission geometry. We have presented the results from nonlinear numerical calculations and compared it with the linear stability analyses for different toy models, and shown that they agree quite well, wherever they are expected to agree.
A necessary condition in order to have fast conversions is the presence of a crossing among the angular spectra of $\nu_e$ and $\bar\nu_e$.
 Another important result is that the backward going modes strongly enhance fast conversions, and make fast conversions possible for a wider range of flux and angular asymmetries.
In particular, rapid flavor turn-overs may occur in time, even if the corresponding spatial evolution is suppressed by the presence of a large matter potential. Considering a moderate hierarchy among the fluxes of different flavors and flavor-dependent $\nu$ angular distributions, as predicted from Garching SN simulations, we show that fast conversion is possible in realistic scenarios. 

The natural region where these effect would show up is just above the core of a SN, where  one expects significant differences in the angular distributions of the different $\nu$ species. 
In multi-dimensional SN simulations presenting the LESA phenomenon, it is quite likely that there will be directions along which the flavor asymmetries among the different species
are moderate at early post-bounce times and allow for these fast conversions. 
If these conversions indeed take place, these would produce a tendency towards flavor equilibrium among the different species. 
If these conversions take place just above the SN core, they may also have a strong impact 
on the revival of the stalled shock-wave by enhanced neutrino reheating and on nucleosynthesis 
of heavy elements~\cite{Dasgupta:2011jf, Pejcha:2011en}.

The possibility of this new form of self-induced flavor conversions however represents a new challenge for the simulation of the SN neutrino signal.  Indeed, in its most general form, the problem involves tracking the neutrino ensemble, including oscillations and
scattering, in an anisotropic and fluctuating environment of an exploding supernova. 
With the current studies we are barely scratching the surface of this tough problem. Certainly, much more work will be needed to achieve a more complete picture of the flavor dynamics of SN neutrinos.

\vspace{-1.0cm}
\section*{Acknowledgments}
We acknowledge useful discussions with Sovan Chakraborty, Thomas Janka, and Georg Raffelt. We thank Thomas Janka for providing us data for the angular distributions from Garching simulations and Francesco Capozzi
for help with processing them.
The work of B.D. is partially supported by the Dept.\,of Science and Technology of the Govt.\,of India through a Ramanujam Fellowship and by the Max-Planck-Gesellschaft through a Max-Planck-Partnergroup. The work of A.M. is supported by the Italian Ministero dell'Istruzione, Universit\`a e Ricerca (MIUR) and Istituto Nazionale di Fisica Nucleare (INFN) through the ``Theoretical Astroparticle Physics'' projects.

\end{document}